\documentclass[10pt,a4paper,onecolumn]{article}
\usepackage{marginnote}
\usepackage{graphicx}
\usepackage{xcolor}
\usepackage{authblk,etoolbox}
\usepackage{titlesec}
\usepackage{calc}
\usepackage{tikz}
\usepackage{hyperref}
\hypersetup{colorlinks,breaklinks,
            urlcolor=[rgb]{0.0, 0.5, 1.0},
            linkcolor=[rgb]{0.0, 0.5, 1.0}}
\usepackage{caption}
\usepackage{tcolorbox}
\usepackage{amssymb,amsmath}
\usepackage{ifxetex,ifluatex}
\usepackage{seqsplit}
\usepackage{fixltx2e} 
\usepackage[
  backend=biber,
]{biblatex}
\bibliography{paper.bib}

\usepackage[top=3.5cm, bottom=3cm, right=1.5cm, left=1.0cm,
            headheight=2.2cm, reversemp, includemp, marginparwidth=4.5cm]{geometry}

\titleformat{\section}
  {\normalfont\sffamily\Large\bfseries}
  {}{0pt}{}
\titleformat{\subsection}
  {\normalfont\sffamily\large\bfseries}
  {}{0pt}{}
\titleformat{\subsubsection}
  {\normalfont\sffamily\bfseries}
  {}{0pt}{}
\titleformat*{\paragraph}
  {\sffamily\normalsize}

\usepackage{fancyhdr}
\makeatletter
\let\ps@plain\ps@fancy
\fancyheadoffset[L]{4.5cm}
\fancyfootoffset[L]{4.5cm}

\definecolor{linky}{rgb}{0.0, 0.5, 1.0}

\newtcolorbox{repobox}
   {colback=red, colframe=red!75!black,
     boxrule=0.5pt, arc=2pt, left=6pt, right=6pt, top=3pt, bottom=3pt}

\newcommand{\ExternalLink}{%
   \tikz[x=1.2ex, y=1.2ex, baseline=-0.05ex]{%
       \begin{scope}[x=1ex, y=1ex]
           \clip (-0.1,-0.1)
               --++ (-0, 1.2)
               --++ (0.6, 0)
               --++ (0, -0.6)
               --++ (0.6, 0)
               --++ (0, -1);
           \path[draw,
               line width = 0.5,
               rounded corners=0.5]
               (0,0) rectangle (1,1);
       \end{scope}
       \path[draw, line width = 0.5] (0.5, 0.5)
           -- (1, 1);
       \path[draw, line width = 0.5] (0.6, 1)
           -- (1, 1) -- (1, 0.6);
       }
   }

\patchcmd{\@maketitle}{center}{flushleft}{}{}
\patchcmd{\@maketitle}{center}{flushleft}{}{}
\patchcmd{\@maketitle}{\LARGE}{\LARGE\sffamily}{}{}
\def\maketitle{{%
  
  \AB@maketitle}}
\makeatletter
\renewcommand\AB@affilsepx{ \protect\Affilfont}
\renewcommand\AB@affilnote[1]{{\bfseries #1}\hspace{3pt}}
\makeatother

\renewcommand\Affilfont{\sffamily\small\mdseries}
\ifnum 0\ifxetex 1\fi\ifluatex 1\fi=0 
  \usepackage[T1]{fontenc}
  \usepackage[utf8]{inputenc}

\else 
  \ifxetex
    \usepackage{mathspec}
  \else
    \usepackage{fontspec}
  \fi
  \defaultfontfeatures{Ligatures=TeX,Scale=MatchLowercase}

\fi
\IfFileExists{upquote.sty}{\usepackage{upquote}}{}
\IfFileExists{microtype.sty}{%
\usepackage{microtype}
\UseMicrotypeSet[protrusion]{basicmath} 
}{}

\usepackage{hyperref}
\hypersetup{unicode=true,
            pdftitle={unyt: Handle, manipulate, and convert data with units in Python},
            pdfborder={0 0 0},
            breaklinks=true}
\usepackage{longtable,booktabs}
\usepackage{graphicx,grffile}
\makeatletter
\def\maxwidth{\ifdim\Gin@nat@width>\linewidth\linewidth\else\Gin@nat@width\fi}
\def\maxheight{\ifdim\Gin@nat@height>\textheight\textheight\else\Gin@nat@height\fi}
\makeatother
\setkeys{Gin}{width=\maxwidth,height=\maxheight,keepaspectratio}
\IfFileExists{parskip.sty}{%
\usepackage{parskip}
}{
\setlength{\parindent}{0pt}
\setlength{\parskip}{6pt plus 2pt minus 1pt}
}
\ifx\paragraph\undefined\else
\let\oldparagraph\paragraph
\renewcommand{\paragraph}[1]{\oldparagraph{#1}\mbox{}}
\fi
\ifx\subparagraph\undefined\else
\let\oldsubparagraph\subparagraph
\renewcommand{\subparagraph}[1]{\oldsubparagraph{#1}\mbox{}}
\fi

\title{unyt: Handle, manipulate, and convert data with units in Python}

        \author[1]{Nathan J. Goldbaum}
          \author[2]{John A. ZuHone}
          \author[1]{Matthew J. Turk}
          \author[1]{Kacper Kowalik}
          \author[2]{Anna L. Rosen}
    
      \affil[1]{National Center for Supercomputing Applications, University of Illinois
at Urbana-Champaign}
      \affil[2]{Harvard-Smithsonian Center for Astrophysics}
  \date{\vspace{-5ex}}

\begin{document}
\maketitle

\marginpar{
  \sffamily\small

  {\bfseries DOI:} \href{https://doi.org/xxx}{\color{linky}{xxx}}

  \vspace{2mm}

  {\bfseries Software}
  \begin{itemize}
    \setlength\itemsep{0em}
    \item \href{xxx}{\color{linky}{Review}} \ExternalLink
    \item \href{https://github.com/yt-project/unyt}{\color{linky}{Repository}} \ExternalLink
    \item \href{xxx}{\color{linky}{Archive}} \ExternalLink
  \end{itemize}

  \vspace{2mm}

  {\bfseries Submitted:} 6 June 2018\\
  {\bfseries Published:} xxx

  \vspace{2mm}
  {\bfseries Licence}\\
  Authors of papers retain copyright and release the work under a Creative Commons Attribution 4.0 International License (\href{http://creativecommons.org/licenses/by/4.0/}{\color{linky}{CC-BY}}).
}

\hypertarget{summary}{%
\section{Summary}\label{summary}}

Software that processes real-world data or that models a physical system
must have some way of managing units. This might be as simple as the
convention that all floating point numbers are understood to be in the
same physical unit system (for example, the SI MKS units system). While
simple approaches like this do work in practice, they also are fraught
with possible error, both by programmers modifying the code who
unintentionally misinterpret the units, and by users of the software who
must take care to supply data in the correct units or who need to infer
the units of data returned by the software. Famously, NASA lost contact
with the Mars Climate Orbiter spacecraft after it crash-landed on the
surface of Mars due to the use of English Imperial units rather than
metric units in the spacecraft control software (Board 1999).

The \texttt{unyt} library is designed both to aid quick calculations at
an interactive python prompt and to be tightly integrated into a larger
Python application or library. The top-level \texttt{unyt} namespace
ships with a large number of predefined units and physical constants to
aid setting up quick calculations without needing to look up unit data
or the value of a physical constant. Using the \texttt{unyt} library as
an interactive calculation aid only requires knowledge of basic Python
syntax and awareness of a few of the methods of the \texttt{unyt\_array}
class - for example, the \texttt{unyt\_array.to()} method to convert
data to a different unit. As the complexity of the usage increases,
\texttt{unyt} provides a number of optional features to aid these cases,
including custom unit registries containing both predefined physical
units as well as user-defined units, built-in output to disk via the
pickle protocol and to HDF5 files using the h5py library (Collette
2013), and round-trip converters for unit objects defined by other
popular Python unit libraries.

Physical units in the \texttt{unyt} class are defined in terms of the
dimensions of the unit, a string representation, and a floating point
scaling to the MKS unit system. Rather than implementing algebra for
unit expressions, we rely on the \texttt{SymPy} symbolic algebra library
(Meurer et al. 2017) to handle symbolic algebraic manipulation. The
\texttt{unyt.Unit} object can represent arbitrary units formed out of
base dimensions in the SI unit system: time, length, mass, temperature,
luminance, and electric current. We currently treat units such as mol
with the seventh SI base dimension, amount of substance, as
dimensionless, although we are open to changing this based on feedback
from users. In addition, \texttt{unyt} supports forming quantities
defined in other unit systems - in particular CGS Gaussian units common
in astrophysics as well as geometrized ``natural'' units common in
relativistic calculations. In addition, \texttt{unyt} ships with a
number of other useful predefined unit systems based, including imperial
units, Planck units, a unit system for calculations in the solar system,
and a ``galactic'' unit system based on the solar mass, kiloparsecs, and
Myr, a convention common in galactic astronomy.

In addition to the \texttt{unyt.Unit} class, \texttt{unyt} also provides
a two subclasses of the NumPy (Oliphant 2006) ndarray (Walt, Colbert,
and Varoquaux 2011), \texttt{unyt.unyt\_array} and
\texttt{unyt.unyt\_quantity} to represent arrays and scalars with units
attached, respectively. The \texttt{unyt} library also provides a
\texttt{unyt.UnitRegistry} class to allow custom systems of units, for
example to track the internal unit system used in a simulation. These
subclasses are tightly integrated with the NumPy ufunc system, which
ensures that algebraic calculations that include data with units
automatically check to make sure the units are consistent, and allow
automatic converting of the final answer of a calculation into a
convenient unit.

We direct readers interested in usage examples and a guide for
integrating \texttt{unyt} into an existing Python application or
workflow to the unyt documentation hosted at
http://unyt.readthedocs.io/en/latest/.

\hypertarget{comparison-with-pint-and-astropy.units}{%
\section{\texorpdfstring{Comparison with \texttt{Pint} and
\texttt{astropy.units}}{Comparison with Pint and astropy.units}}\label{comparison-with-pint-and-astropy.units}}

The scientific Python ecosystem has a long history of efforts to develop
a library to handle unit conversions and enforce unit consistency. For a
relatively recent review of these efforts, see (Bekolay 2013). While we
won't exhaustively cover extant Python libraries for handling units in
this paper, we will focus on \texttt{Pint} (Grecco 2018) and
\texttt{astropy.units} (The Astropy Collaboration et al. 2018), which
both provide a robust implementation of an array container with units
attached and are commonly used in research software projects. At time of
writing a GitHub search for \texttt{import\ astropy.units} returns
approximately 10,500 results and a search for \texttt{import\ pint}
returns approximately 1,500 results.

While \texttt{unyt} provides functionality that overlaps with
\texttt{astropy.units} and \texttt{Pint}, there are important
differences which we elaborate on below. In addition, it is worth noting
that all three codebases had origins at roughly the same time period.
\texttt{Pint} initially began development in 2012 according to the git
repository logs. Likewise \texttt{astropy.units} began development in
2012 and was released as part of \texttt{astropy\ 0.2} in 2013, although
the initial implementation was adapted from the \texttt{pynbody} library
(Pontzen et al. 2013), which started its units implementation in 2010
according to the git repository logs. In the case of \texttt{unyt}, it
originated via the \texttt{dimensionful} library (Stark 2012) in 2012.
Later, \texttt{dimensionful} was elaborated on and improved to become
\texttt{yt.units}, the unit system for the \texttt{yt} library (Turk et
al. 2011) at a \texttt{yt} developer workshop in 2013 and was
subsequently released as part of \texttt{yt\ 3.0} in 2014. One of the
design goals for the \texttt{yt} unit system was the ability to
dynamically define ``code'' units (e.g.~units internal to data loaded by
yt) as well as units that depend on details of the dataset - in
particular cosmological comoving units and the ``little \(h\)'' factor,
used to parameterize the Hubble constant in cosmology calculations
(Croton 2013). For cosmology simulations in particular, comparing data
with different unit systems can be tricky because one might want to use
data from multiple outputs in a time series, with each output having a
different mapping from internal units to physical units. This despite
the fact that each output in the time series represents the same
physical system and common workflows involve combining data from
multiple outputs. This requirement to manage complex custom units and
interoperate between custom unit systems drove the \texttt{yt} community
to independently develop a custom unit system solution. We have decided
to repackage and improve \texttt{yt.units} in the form of \texttt{unyt}
to both make it easier to work on and improve the unit system and
encourage use of the unit system for scientific python users who do not
want to install a heavy-weight dependency like \texttt{yt}.

Below we present a table comparing \texttt{unyt} with
\texttt{astropy.units} and \texttt{Pint}. Estimates for lines of code in
the library were generated using the \texttt{cloc} tool (Danial 2018);
blank and comment lines are excluded from the estimate. Test coverage
was estimated using the \texttt{coveralls} output for \texttt{Pint} and
\texttt{astropy.units} and using the \texttt{codecov.io} output for
\texttt{unyt}.

\begin{longtable}[]{@{}llll@{}}
\toprule
Library & \texttt{unyt} & \texttt{astropy.units} &
\texttt{Pint}\tabularnewline
\midrule
\endhead
Lines of code & 5128 & 10163 & 8908\tabularnewline
Lines of code excluding tests & 3195 & 5504 & 4499\tabularnewline
Test Coverage & 99.91\% & 93.63\% & 77.44\%\tabularnewline
\bottomrule
\end{longtable}

We offer lines of code as a very rough estimate for the ``hackability''
of the codebase. In general, smaller codebases with higher test coverage
have fewer defects (Lipow 1982; Koru, Zhang, and Liu 2007; Gopinath,
Jensen, and Groce 2014). This comparison is somewhat unfair in favor of
\texttt{unyt} in that \texttt{astropy.units} only depends on NumPy and
\texttt{Pint} has no dependencies, while \texttt{unyt} depends on both
\texttt{SymPy} and NumPy. Much of the reduction in the size of the
\texttt{unyt} library can be attributed to offloading the handling of
algebra to \texttt{SymPy} rather than needing to implement the algebra
of unit symbols directly in \texttt{unyt}. For potential users who are
wary of adding \texttt{SymPy} as a dependency, that might argue in favor
of using \texttt{Pint} in favor of \texttt{unyt}.

\hypertarget{astropy.units}{%
\subsection{\texorpdfstring{\texttt{astropy.units}}{astropy.units}}\label{astropy.units}}

The \texttt{astropy.units} subpackage provides a \texttt{PrefixUnit}
class, a \texttt{Quantity} class that represents both scalar and array
data with attached units, and a large number of predefined unit symbols.
The preferred way to create \texttt{Quantity} instances is via
multiplication with a \texttt{PrefixUnit} instance. Similar to
\texttt{unyt}, the \texttt{Quantity} class is implemented via a subclass
of the NumPy \texttt{ndarray} class. Indeed, in many ways the everyday
usage patterns of \texttt{astropy.units} and \texttt{unyt} are similar,
although \texttt{unyt} is not quite a drop-in replacement for
\texttt{astropy.units} as there are some API differences. The main
functional difference between \texttt{astropy.units} and \texttt{unyt}
is that \texttt{astropy.units} is a subpackage of the larger
\texttt{astropy} package. This means that depending on
\texttt{astropy.units} requires installing a large collection of
astronomically focused software included in the \texttt{astropy}
package, including a substantial amount of compiled C code. This
presents a barrier to usage for potential users of
\texttt{astropy.units} who are not astronomers or do not need the
observational astronomy capabilities provided by \texttt{astropy}.

\hypertarget{pint}{%
\subsection{\texorpdfstring{\texttt{Pint}}{Pint}}\label{pint}}

The \texttt{Pint} package provides a different API for accessing units
compared with \texttt{unyt} and \texttt{astropy.units}. Rather than
making units immediately importable from the \texttt{Pint} namespace,
\texttt{Pint} instead requires users to instantiate a
\texttt{UnitRegistry} instance (unrelated to the
\texttt{unyt.UnitRegistry} class), which in turn has \texttt{Unit}
instances as attributes. Just like with \texttt{unyt} and
\texttt{astropy.units}, creating a \texttt{Quantity} instance requires
multiplying an array or scalar by a \texttt{Unit} instance. Exposing the
\texttt{UnitRegistry} directly to all users like this does force users
of the library to think about which system of units they are working
with, which may be beneficial in some cases, however it also means that
users have a bit of extra cognitive overhead they need to deal with
every time they use \texttt{Pint}.

\begin{figure}
\centering
\includegraphics{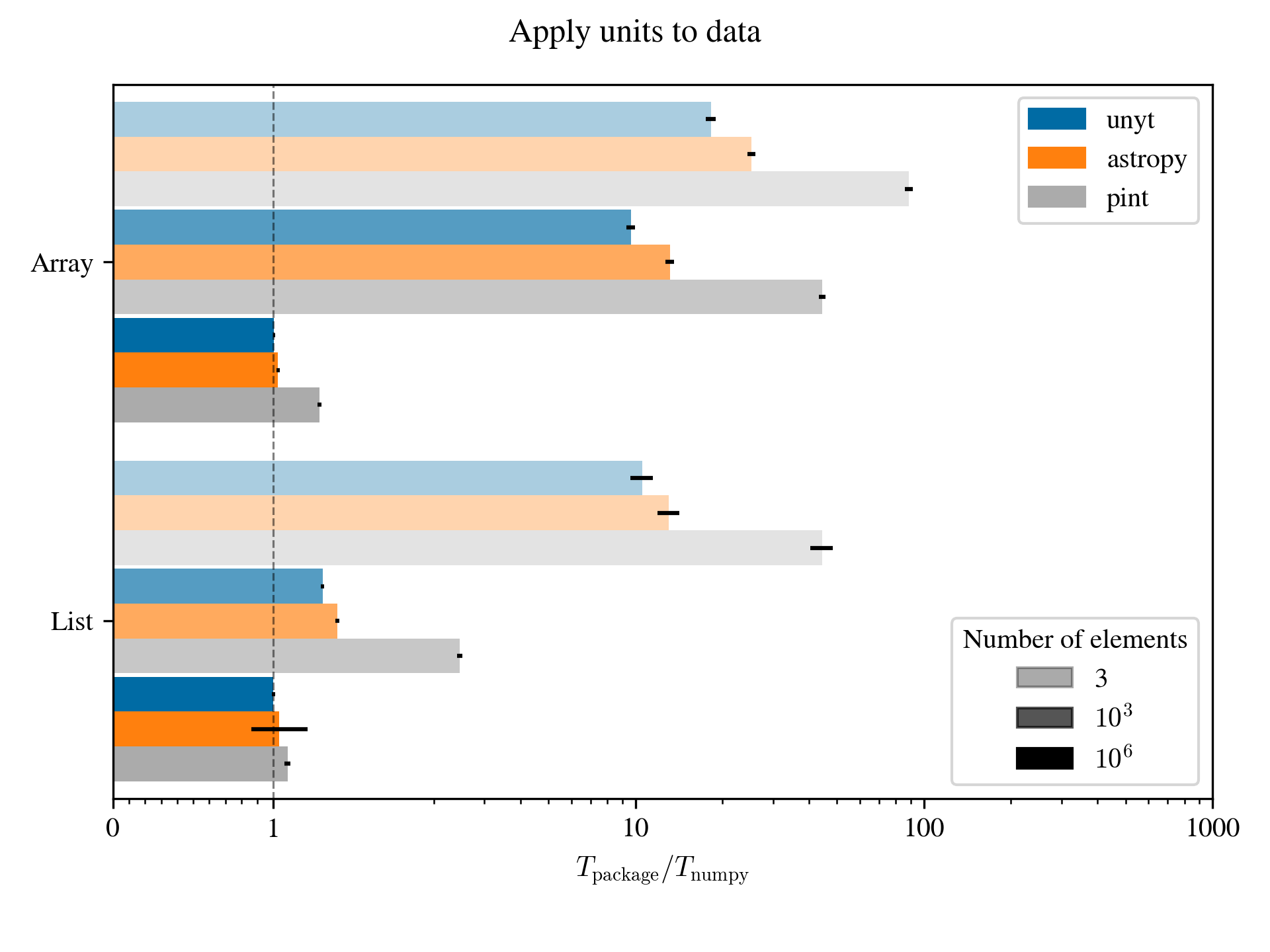}
\caption{A benchmark comparing the ratio of the time to apply units to
lists and NumPy \texttt{ndarray} instances to the time to interpret the
same list or \texttt{ndarray} to an \texttt{ndarray}. This ratio,
\(T_{\rm package}/T_{\rm numpy}\), corresponds to the overhead of
converting data to work with one of the three packages. Values close to
unity correspond to zero or negligible overhead, while values larger
than unity correspond to measureable overhead. Optimally all values
would be near zero. In practice, applying units to small arrays incurs
substantial overhead. Each test is shown for three different sizes of
input data, including inputs with size 3, 1,000, and 1,000,000. The
black lines at the top of the bars indicate the sample standard
deviation. The \(T_{\rm numpy}\) time is calculated by benchmarking the
time to perform \texttt{np.asarray(data)} where \texttt{data} is either
a \texttt{list} or an \texttt{ndarray}.}
\end{figure}

In addition, the \texttt{Quantity} class provided by \texttt{Pint} is
not a subclass of NumPy's ndarray. Instead, it is a wrapper around an
internal \texttt{ndarray} buffer. This simplifies the implementation of
\texttt{Pint} by avoiding the somewhat arcane process for creating an
ndarray subclass, although the \texttt{Pint} \texttt{Quantity} class
must also be careful to emulate the full NumPy \texttt{ndarray} API so
that it can be a drop-in replacement for \texttt{ndarray}.

Finally, in comparing the output of our benchmarks of \texttt{Pint},
\texttt{astropy.units}, and \texttt{unyt}, we found that in-place
operations making use of a NumPy \texttt{ufunc} will unexpectedly strip
units in \texttt{Pint}. For example, if \texttt{a} and \texttt{b} are
\texttt{Pint} \texttt{Quantity} instances,
\texttt{np.add(a,\ b,\ out=out))} will operate on \texttt{a} and
\texttt{b} as if neither have units attached. Interestingly, without the
\texttt{out} keyword, \texttt{Pint} does get the correct answer, so it
is possible that this is a bug in \texttt{Pint}, and we have reported it
as such upstream (see https://github.com/hgrecco/pint/issues/644).

\hypertarget{performance-comparison}{%
\subsection{Performance Comparison}\label{performance-comparison}}

Checking units will always add some overhead over using hard-coded unit
conversion factors. Thus a library that is entrusted with checking units
in an application should incur the minimum possible overhead to avoid
triggering performance regressions after integrating unit checking into
an application. Optimally, a unit library will add zero overhead
regardless of the size of the array. In practice that is not the case
for any of the three libraries under consideration, and there is a
minimum array size above which the overhead of doing a mathematical
operation exceeds the overhead of checking units. It is thus worth
benchmarking unit libraries in a fair manner, comparing with the same
operation implemented using plain NumPy.

\begin{figure}
\centering
\includegraphics{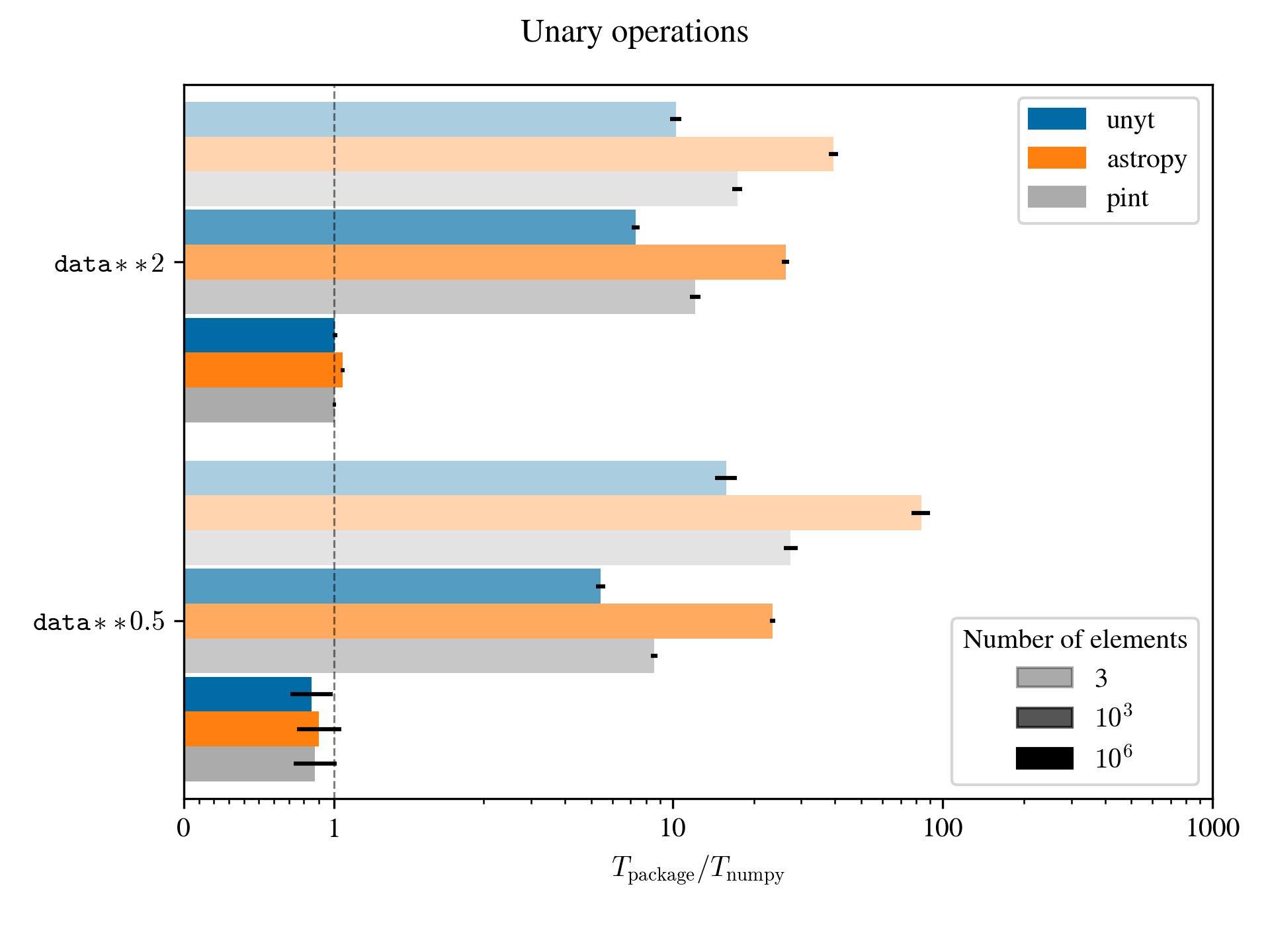}
\caption{A benchmark comparing the time to square an array and to take
the square root of an array. See Figure 1 for a detailed explanation of
the plot style.}
\end{figure}

Here we present such a benchmark. We made use of the \texttt{perf}
(Stinner 2018) Python benchmarking tool, which not only provides
facilities for establishing the statistical significance of a benchmark
run, but also can tune a linux system to turn off operating system and
hardware features like CPU throttling that might introduce variance in a
benchmark. We made use of a Dell Latitude E7270 laptop equipped with an
Intel i5-6300U CPU clocked at 2.4 Ghz. The testing environment was based
on \texttt{Python\ 3.6.3} and had \texttt{NumPy\ 1.14.2},
\texttt{sympy\ 1.1.1}, \texttt{fastcache\ 1.0.2},
\texttt{Astropy\ 3.0.1}, and \texttt{Pint\ 0.8.1} installed.
\texttt{fastcache} (Brady 2017) is an optional dependency of
\texttt{SymPy} that provides an optimized LRU cache implemented in C
that can substantially speed up \texttt{SymPy}. The system was
instrumented using \texttt{perf\ system\ tune} to turn off CPU features
that might interfere with stable benchmarks. We did not make any
boot-time Linux kernel parameter changes.

\begin{figure}
\centering
\includegraphics{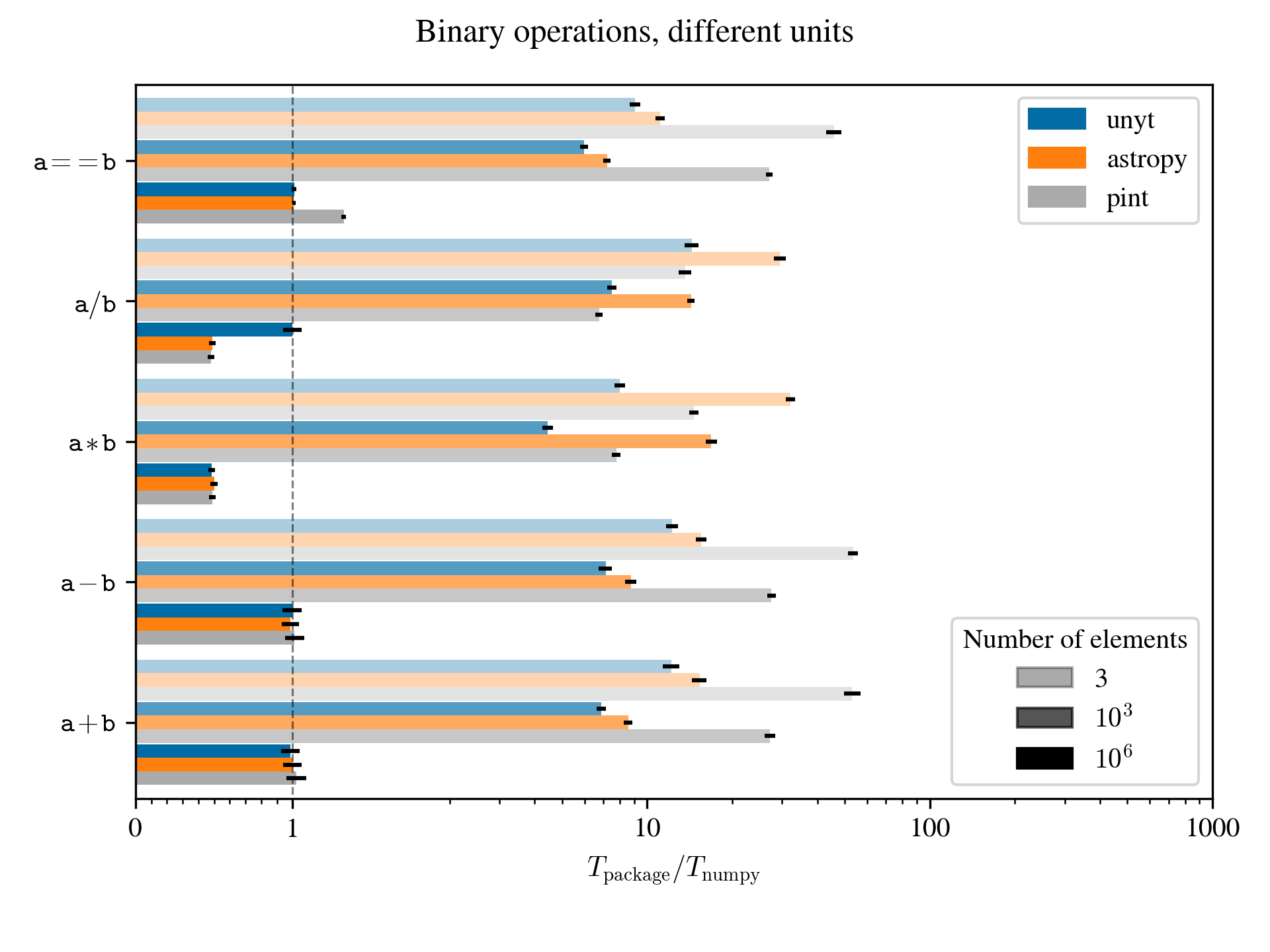}
\caption{A benchmark comparing the time to perform various binary
arithmetic operations on input operans that have different but
dimensionallty compatible units. See Figure 1 for a detailed explanation
of the plot style.}
\end{figure}

For each of the benchmarks below, we show the ratio of the time to
perform an operation with one of \texttt{unyt}, \texttt{Pint}, and
\texttt{astopy.units}, \(T_{\rm package}\), to the time it takes for
NumPy to perform the equivalent operation, \(T_{\rm numpy}\). For
example, for the comparison of the performance of \texttt{np.add(a,\ b)}
where \texttt{a} and \texttt{b} have different units with the same
dimension, the corresponding benchmark to generate \(T_{\rm numpy}\)
would use the code \texttt{np.add(a,\ c*b)} where \texttt{a} and
\texttt{b} would be \texttt{ndarray} instances and \texttt{c} would be
the floating point conversion factor between the units of \texttt{a} and
\texttt{b}. Much of the time in \(T_{\rm package}\) relative to
\(T_{\rm numpy}\) is spent in the respective packages calculating the
appropriate conversion factor \texttt{c}. Thus the comparisons below
depict very directly the overhead for using a unit library over an
equivalent operation that uses hard-coded unit-conversion factors.

\begin{figure}
\centering
\includegraphics{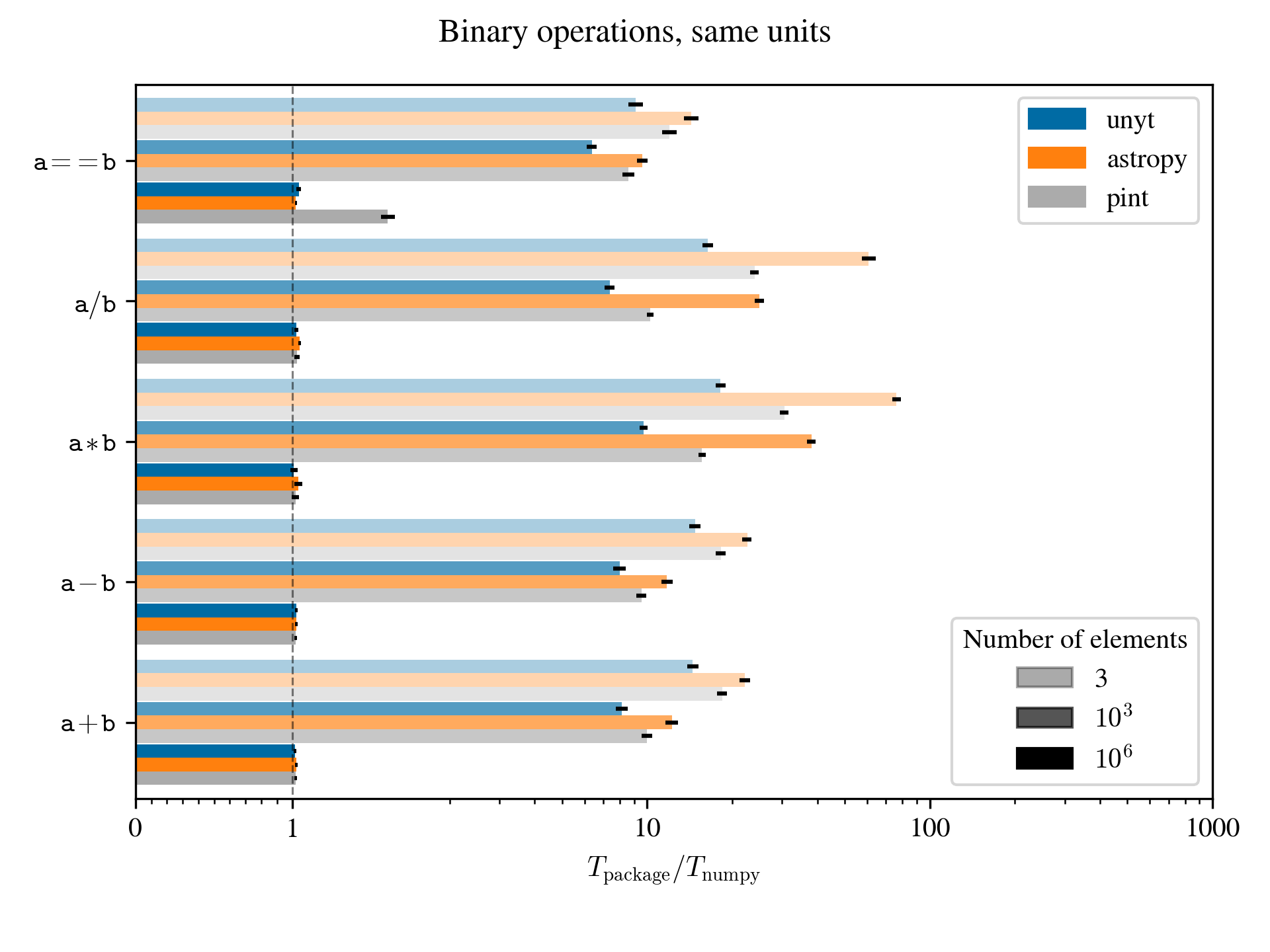}
\caption{A benchmark comparing the time to perform various binary
arithmetic operations on input operands that have the same units. See
Figure 1 for a detailed explanation of the plot style.}
\end{figure}

\hypertarget{applying-units-to-data}{%
\subsubsection{Applying units to data}\label{applying-units-to-data}}

In Figure 1 we plot the overhead for applying units to data, showing
both Python lists and NumPy \texttt{ndarray} instances as the input to
apply data to. Since all three libraries eventually convert input data
to a NumPy \texttt{ndarray}, the comparison with array inputs more
explicitly shows the overhead for \emph{just} applying units to data.
When applying units to a list, all three libraries as well as NumPy need
to first copy the contents of the list into a NumPy array or a subclass
of \texttt{ndarray}. This explains why the overhead is systematically
lower when starting with a list.

\begin{figure}
\centering
\includegraphics{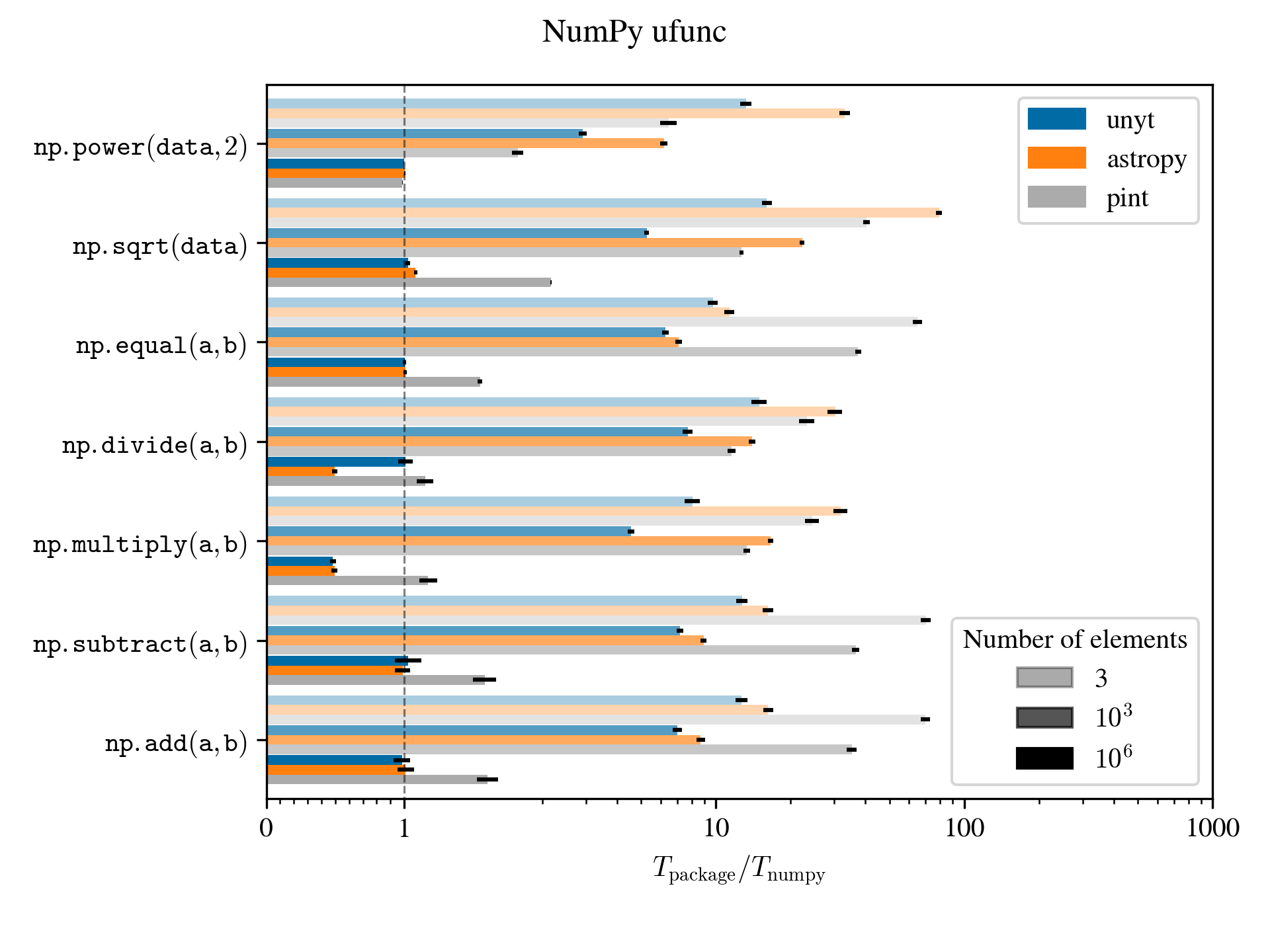}
\caption{A benchmark comparing the overhead for computing various NumPy
\texttt{ufunc} operations. The operands of all binary \texttt{ufuncs}
have the same units. See Figure 1 for a detailed explanation of the plot
style.}
\end{figure}

In all cases, \texttt{unyt} either is fastest by a statistically
significant margin, or ties with \texttt{astropy}. Even for large input
arrays, \texttt{Pint} still has statistically significant overhead,
while both \texttt{unyt} and \texttt{astropy.units} have negligible
overhead once the input array size reaches \(10^6\) elements.

\hypertarget{unary-arithmetic-operations}{%
\subsubsection{Unary arithmetic
operations}\label{unary-arithmetic-operations}}

Expressions involving powers of data with units, including integer and
fractional powers, are very common in the physical sciences. It is
therefore very important for a library that handles units to be able to
track this case in a performant way. In Figure 2 we present a benchmark
comparing \texttt{Pint}, \texttt{unyt}, and \texttt{astropy.units} for
the squaring and square root operation. In all cases, \texttt{unyt} has
the lowest overhead, with \texttt{Pint} coming in second, and
\texttt{astropy.units} trailing. Note that the y-axis is plotted on a
log scale, so \texttt{astropy} is as much as 4 times slower than
\texttt{unyt} for these operations.

\hypertarget{binary-arithmetic-operations}{%
\subsubsection{Binary arithmetic
operations}\label{binary-arithmetic-operations}}

Binary operations form the core of arithmetic. It is vital for a library
that handles unit manipulation to both transparently convert units when
necessary and to ensure that expressions involving quantities with units
are dimensionally consistent. In Figure 3 and 4 we present benchmarks
for binary arithmetic expressions, both with input data that has the
same units and with input data with different units but the same
dimensions. In most cases, \texttt{unyt} has less overhead than both
\texttt{astropy} and \texttt{Pint}, although there are a few anomalies
that are worth explaining in more detail. For comparison operations,
\texttt{Pint} exhibits a slowdown even on large input arrays. This is
not present for other binary operations, so it is possible that this
overhead might be eliminated with a code change in \texttt{Pint}. For
multiplication on large arrays, all three libraries have measured
overhead of \(\sim 0.5\) than that of the ``equivalent'' Numpy
operation. That is because all three libraries produce results with
units given by the product of the input units. That is, there is no need
to multiply the result of the multiplication operation by an additional
constant, while a pure NumPy implementation would need to multiply the
result by a constant to ensure both operands of the operation are in the
same units. Finally, for division, both \texttt{Pint} and
\texttt{astropy.units} exhibit the same behavior as for multiplication,
and for similar reasons: the result of the division operation is output
with units given by the ratio of the input units. On the other hand,
\texttt{unyt} will automatically cancel the dimensionally compatible
units in the ratio and return a result with dimensionless units.

\begin{figure}
\centering
\includegraphics{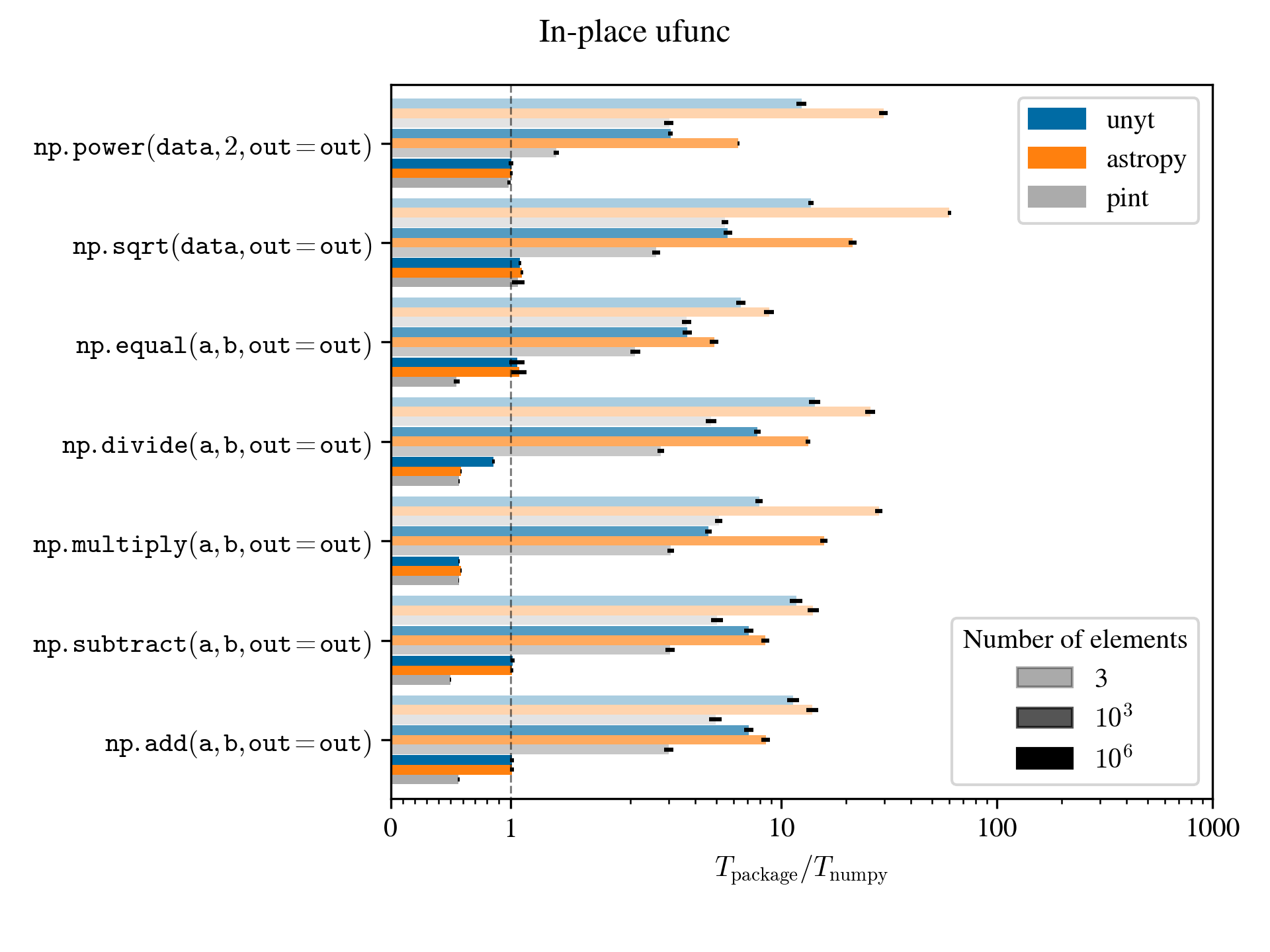}
\caption{The same as Figure 5, but with in-place \texttt{ufunc}
operations. See Figure 1 for a detailed explanation of the plot style.}
\end{figure}

\hypertarget{numpy-ufunc-performance}{%
\subsubsection{\texorpdfstring{NumPy \texttt{ufunc}
performance}{NumPy ufunc performance}}\label{numpy-ufunc-performance}}

Lastly, In Figures 4 and 5, we present benchmarks of NumPy
\texttt{ufunc} operations. A NumPy \texttt{ufunc} is a fast C
implementation of a basic mathematical operation. This includes
arithmetic operators as well as trigonometric and special functions. By
using a \texttt{ufunc} directly, one bypasses the Python object protocol
and short-circuits directly to the low-level NumPy math kernels. We show
both directly using the NumPy \texttt{ufunc} operators (Figure 4) as
well as using the same operators with a pre-allocated output array to
benchmark in-place operations.

As for the other benchmarks, \texttt{unyt} tends to have the lowest
amount of overhead, although there are some significant exceptions. For
\texttt{np.power}, \texttt{Pint} has the lowest overhead, except for
very large input arrays, where the overhead for all three libraries is
negligible. On the other hand, for \texttt{np.sqrt}, \texttt{np.equal},
\texttt{np.add}, and \texttt{np.subtract}, \texttt{Pint} still has
statistically significant overhead for large input arrays. Finally, for
the in-place \texttt{ufunc} comparison, \texttt{Pint} has the lowest
overhead for all operations. However, as discussed above, this is
because of a bug in \texttt{Pint} which causes the library to ignore
units when calling a \texttt{ufunc} with the \texttt{out} keyword set.

\hypertarget{conclusions}{%
\section{Conclusions}\label{conclusions}}

In this paper we present the \texttt{unyt} library, giving background on
the reasons for its existence and some historical context for its
origin. We also present a set of benchmarks for common arithmetic
operations, comparing the performance of \texttt{unyt} with
\texttt{Pint} and \texttt{astropy.units}. In general, we find that
\texttt{unyt} either outperforms or matches the performance
\texttt{astropy.units} and \texttt{Pint}, depending on the operation and
size of the input data. We also demonstrate that the \texttt{unyt}
library constitutes a smaller codebase with higher test coverage than
both \texttt{Pint} and \texttt{astropy.units}.

\hypertarget{acknowledgements}{%
\section{Acknowledgements}\label{acknowledgements}}

NJG would like to thank Brandon Carswell and Alex Farthing of the NCSA
IT staff for providing a laptop with Linux installed for the performance
benchmark. This work was supported by NSF grant OAC-1663914 (NJG, MJT),
by the Gordon and Betty Moore Foundation's Data-Driven Discovery
Initiative through Grant GBMF4561 (MJT) and by NASA through Einstein
Postdoctoral Fellowship grant number PF7-180166 awarded by the Chandra
X-ray Center, which is operated by the Smithsonian Astrophysical
Observatory for NASA under contract NAS8-03060 (ALR).

\hypertarget{references}{%
\section*{References}\label{references}}
\addcontentsline{toc}{section}{References}

\hypertarget{refs}{}
\leavevmode\hypertarget{ref-bekolay2013}{}%
Bekolay, Trevor. 2013. ``A Comprehensive Look at Representing Physical
Quantities in Python.''
\url{https://www.youtube.com/watch?v=N-edLdxiM40}.

\leavevmode\hypertarget{ref-nasa1999}{}%
Board, Mars Climate Orbiter Mishap Investigation. 1999. \emph{Mars
Climate Orbiter Mishap Investigation Board: Phase I Report}. Jet
Propulsion Laboratory.
\url{https://books.google.com/books?id=4OMIHQAACAAJ}.

\leavevmode\hypertarget{ref-fastcache}{}%
Brady, Peter. 2017. ``Fastcache.'' \emph{GitHub Repository}.
\url{https://github.com/pbrady/fastcache}; GitHub.

\leavevmode\hypertarget{ref-h5py}{}%
Collette, Andrew. 2013. \emph{Python and Hdf5}. O'Reilly.

\leavevmode\hypertarget{ref-croton2013}{}%
Croton, D. J. 2013. ``Damn You, Little h! (Or, Real-World Applications
of the Hubble Constant Using Observed and Simulated Data)'' 30
(October): e052. \url{https://doi.org/10.1017/pasa.2013.31}.

\leavevmode\hypertarget{ref-cloc}{}%
Danial, Al. 2018. ``Cloc.'' \emph{GitHub Repository}.
\url{https://github.com/AlDanial/cloc}; GitHub.

\leavevmode\hypertarget{ref-Gopinath2014}{}%
Gopinath, Rahul, Carlos Jensen, and Alex Groce. 2014. ``Code Coverage
for Suite Evaluation by Developers.'' In \emph{Proceedings of the 36th
International Conference on Software Engineering}, 72--82. ICSE 2014.
New York, NY, USA: ACM. \url{https://doi.org/10.1145/2568225.2568278}.

\leavevmode\hypertarget{ref-Pint}{}%
Grecco, Hernan E. 2018. ``Pint.'' \emph{GitHub Repository}.
\url{https://github.com/hgrecco/pint}; GitHub.

\leavevmode\hypertarget{ref-Koru2007}{}%
Koru, A. G., D. Zhang, and H. Liu. 2007. ``Modeling the Effect of Size
on Defect Proneness for Open-Source Software.'' In \emph{Predictor
Models in Software Engineering, 2007. PROMISE'07: ICSE Workshops 2007.
International Workshop on}, 10--10.
\url{https://doi.org/10.1109/PROMISE.2007.9}.

\leavevmode\hypertarget{ref-Lipow1982}{}%
Lipow, M. 1982. ``Number of Faults Per Line of Code.'' \emph{IEEE Trans.
Softw. Eng.} 8 (4). Piscataway, NJ, USA: IEEE Press: 437--39.
\url{https://doi.org/10.1109/TSE.1982.235579}.

\leavevmode\hypertarget{ref-SymPy}{}%
Meurer, Aaron, Christopher P. Smith, Mateusz Paprocki, Ondřej Čertík,
Sergey B. Kirpichev, Matthew Rocklin, AMiT Kumar, et al. 2017. ``SymPy:
Symbolic Computing in Python.'' \emph{PeerJ Computer Science} 3
(January): e103. \url{https://doi.org/10.7717/peerj-cs.103}.

\leavevmode\hypertarget{ref-NumPy}{}%
Oliphant, Travis E. 2006. \emph{A Guide to Numpy}. Trelgol Publishing.

\leavevmode\hypertarget{ref-pynbody}{}%
Pontzen, A., R. Roškar, G. S. Stinson, R. Woods, D. M. Reed, J. Coles,
and T. R. Quinn. 2013. ``pynbody: Astrophysics Simulation Analysis for
Python.''

\leavevmode\hypertarget{ref-dimensionful}{}%
Stark, Casey W. 2012. ``Dimensionful.'' \emph{GitHub Repository}.
\url{https://github.com/caseywstark/dimensionful}; GitHub.

\leavevmode\hypertarget{ref-perf}{}%
Stinner, Victor. 2018. ``Perf.'' \emph{GitHub Repository}.
\url{https://github.com/vstinner/perf}; GitHub.

\leavevmode\hypertarget{ref-astropy}{}%
The Astropy Collaboration, A. M. Price-Whelan, B. M. Sipőcz, H. M.
Günther, P. L. Lim, S. M. Crawford, S. Conseil, et al. 2018. ``The
Astropy Project: Building an inclusive, open-science project and status
of the v2.0 core package.'' \emph{ArXiv E-Prints}, January.

\leavevmode\hypertarget{ref-yt}{}%
Turk, M. J., B. D. Smith, J. S. Oishi, S. Skory, S. W. Skillman, T.
Abel, and M. L. Norman. 2011. ``yt: A Multi-code Analysis Toolkit for
Astrophysical Simulation Data.'' \emph{The Astrophysical Journal
Supplement Series} 192 (January): 9.
\url{https://doi.org/10.1088/0067-0049/192/1/9}.

\leavevmode\hypertarget{ref-vanderwalt2011}{}%
Walt, S. van der, S. C. Colbert, and G. Varoquaux. 2011. ``The Numpy
Array: A Structure for Efficient Numerical Computation.''
\emph{Computing in Science Engineering} 13 (2): 22--30.
\url{https://doi.org/10.1109/MCSE.2011.37}.

\end{document}